\documentclass[11pt,draftcls,onecolumn]{article}

\usepackage{makeidx}
\usepackage{amsmath}
\usepackage{epsfig}
\usepackage{amssymb}
\usepackage{textcomp}
\usepackage{pstricks,pst-text}
\usepackage{algorithm}
\usepackage{algorithmic}
\usepackage{datetime}
\usepackage{setspace}
\usepackage{stfloats}
\usepackage{subcaption}
\usepackage{cases}
\usepackage{cite}
\usepackage{watermark}

\usepackage{amsmath,graphicx}

\newcommand{\mbf}{\mathbf}

\newcommand{\bzero}{\mbf{0}}

\def\b1{{\mathbf 1}}

\def\bSigma{{\mbox{\boldmath{$\Sigma$}}}}
\def\cSigma{\emph{\mbox{\boldmath{$\Sigma$}}}}

\def\btheta{{\mbox{\boldmath{$\theta$}}}}

\def\bEta{{\mbox{\boldmath{$\eta$}}}}

\def\iid{{i.i.d.\ }}

\newcommand{\balpha}{\boldsymbol\alpha}
\newcommand{\bbeta}{\boldsymbol\beta}
\newcommand{\bomega}{\boldsymbol\omega}
\newcommand{\bphi}{\boldsymbol\phi}

\def\bepsilon{{\mbox{\boldmath{$\epsilon$}}}}
\def\bgamma{{\mbox{\boldmath{$\gamma$}}}}

\def\blambda{{\mbox{\boldmath{$\lambda$}}}}

\def\bphi{{\mbox{\boldmath{$\phi$}}}}

\def\bSigma{{\mbox{\boldmath{$\Sigma$}}}}

\def\bdelta{{\mbox{\boldmath{$\delta$}}}}

\newcommand{\cA}{\ensuremath{\mathcal{A}}}
\newcommand{\cB}{\ensuremath{\mathcal{B}}}

\newcommand{\cD}{\ensuremath{\mathcal{D}}}

\newcommand{\cG}{\ensuremath{\mathcal{G}}}

\newcommand{\cN}{\ensuremath{\mathcal{N}}}

\newcommand{\diag}{\mbox{\rm diag}}

\def\eg{{e.g.,\ }}
\def\ie{{i.e.,\ }}

\newcommand{\betab}{\begin{tabbing}}
\newcommand{\entab}{\end{tabbing}}
\newcommand{\beitem}{\begin{itemize}}
\newcommand{\enitem}{\end{itemize}}
\newcommand{\bea}{\begin{array}}
\newcommand{\ena}{\end{array}}
\newcommand{\beq}{\begin{equation}}
\newcommand{\enq}{\end{equation}}
\newcommand{\beqa}{\begin{eqnarray}}
\newcommand{\enqa}{\end{eqnarray}}
\newcommand{\beqan}{\begin{eqnarray*}}
\newcommand{\enqan}{\end{eqnarray*}}
\newcommand{\beenum}{\begin{enumerate}}
\newcommand{\enenum}{\end{enumerate}}
\newcommand{\DL}{\begin{dashlist}}
\newcommand{\DLE}{\end{dashlist}}

\newcommand{\bc}{{\ensuremath{\mathbf{c}}}}
\newcommand{\bd}{{\ensuremath{\mathbf{d}}}}
\newcommand{\be}{{\ensuremath{\mathbf{e}}}}

\newcommand{\bq}{{\ensuremath{\mathbf{q}}}}
\newcommand{\br}{{\ensuremath{\mathbf{r}}}}

\newcommand{\bt}{{\ensuremath{\mathbf{t}}}}

\newcommand{\brdot}{{\ensuremath{\mathbf{\dot{\br}}}}}
\newcommand{\brrdot}{{\ensuremath{\mathbf{\ddot{\br}}}}}

\newcommand{\bA}{{\ensuremath{\mathbf{A}}}}

\newcommand{\bC}{{\ensuremath{\mathbf{C}}}}

\newcommand{\bE}{{\ensuremath{\mathbf{E}}}}
\newcommand{\bF}{{\ensuremath{\mathbf{F}}}}

\newcommand{\bI}{{\ensuremath{\mathbf{I}}}}
\newcommand{\bJ}{{\ensuremath{\mathbf{J}}}}

\newcommand{\bT}{{\ensuremath{\mathbf{T}}}}
\newcommand{\bU}{{\ensuremath{\mathbf{U}}}}

\newcommand{\firstAuthor}       {Raj~Thilak~Rajan}
\newcommand{\secondAuthor}      {Alle-Jan~van~der~Veen}

\newcommand{\theTitle}	{Joint non-linear ranging and affine synchronization \\ basis for a network of mobile nodes}

\title{\theTitle}
\author{\firstAuthor$^{*, \dagger}$ and \secondAuthor$^{\dagger}$
\thanks{This research was funded in part by the STW OLFAR project (Contract Number: 10556) within the ASSYS perspectief program.}}

\newcounter{eqnCounter2}

\newcommand{\EEPLS}     {$\text{E}^2\text{PLS}$}
\newcommand{\EEGLS}     {$\text{E}^2\text{GLS}$}
\newcommand{\EExtended} {$(\text{Extended})^2$}

\begin{document}
\watermark{\small{R.T.Rajan, A.-J.van der Veen}}
\date{}
\maketitle

\begin{abstract} Synchronization and localization are critical challenges for the coherent functioning of a wireless network of mobile nodes. In this paper,  a novel joint non-linear range and affine time model is presented based on two way time stamp exchanges, extending an existing affine time-range model. For a pair of nodes, a closed form pairwise least squares solution is proposed for estimating pairwise range parameters, namely relative range, range rate and rate of range rate between the nodes, in addition to estimating the clock skews and the clock offsets. Extending these pair wise solutions to network wide ranging and clock synchronization, we present a central data fusion based global least squares solution. Furthermore, a new Constrained Cramer Rao Bound (CCRB) is derived for the joint time-range model and the proposed algorithms are shown to approach the theoretical limits.\vspace{0.5mm}
\end{abstract}

\section{Introduction} \label{sec:intro} The coherent functioning of wireless networks relies heavily on time synchronization among nodes \cite{sundaram05}. All nodes in a network are equipped with independent clock oscillators, which must be synchronized to a global reference, to facilitate accurate time stamping of data and synchronized communication of processed information. Furthermore, when nodes are mobile and/or arbitrarily deployed in the field, then position estimation is often equally critical as time synchronization \cite{patwari05}. The intermediate distances between all the nodes in the network is one of the key inputs for almost all localization techniques.

Among various potential applications, our key motivation is the Orbiting Low Frequency Antennas for Radio astronomy (OLFAR) \cite{rajanAero11}, which aims to design and develop a detailed system concept for an interferometric array ($\ge$ 10) of identical, scalable and autonomous satellites in space to be used as a scientific instrument for ultra low frequency observations. Due to its distant deployment location, far from the earth orbiting global positioning systems, and the large number of satellites, autonomous network synchronization and localization is one of the key challenges in OLFAR.

For a fixed network of immobile nodes capable of two way communication \cite{ieee07}, various least squares solutions are prevalent for clock synchronization, which model each node clock as a first order polynomial and, subsequently estimate clock skews and clock offsets \cite{wu11}. As an extension, the Global least Squares (GLS) estimator was presented in \cite{rajanCAMSAP11} to estimate the clock parameters along the pairwise distances between all the nodes in the network. A step further, for a network with mobile nodes, an affine time-range model was proposed in \cite{rajanICASSP12}, which approximates the time varying pairwise distance to the first order. Using this model, an Extended Global Least Squares (EGLS) solution was presented to estimate the clock skews, offsets and in addition the ranges and range rates of the network. \emph{However, the pairwise Euclidean distance between a cluster of mobile nodes are always non-linear}. Hence, as an extension of the affine time-range model, we propose a novel non-linear range model in conjunction with an affine clock model. For a pair of mobile nodes capable of two way communication, we present \text{(Extended)$^2$} Pairwise Least Squares (\EEPLS) solution to estimate clock parameters upto first order and range parameters upto the second order. In addition, a centralized \text{(Extended)$^2$} Global Least Squares (\EEGLS) is proposed for estimating clock and range parameters across the network. \section{Joint time range model} \label{sec:timeRange}  \subsection{Time} \label{sec:affineTime} Consider a network of $N$ nodes equipped with independent clock oscillators which, under ideal
conditions, are synchronized to the global time. However,
in reality, due to various oscillator imperfections and
environment conditions the clocks vary independently. Let
$t_i$ be the local time at node $i$, then its divergence
from the ideal \emph{true} time $t$ is to first order given by
the affine clock model \cite{wu11,rajanICASSP12,rajanCAMSAP11},
\begin{eqnarray}
  t_{i} = \omega_{i}t+ \phi_i \quad \Leftrightarrow \quad\ t= \alpha_{i}t_i +\beta_{i}
  \label{eq:affineTime}
\end{eqnarray} where $\omega_{i} \in \mathbb{R}_+$ and $\phi_i\in
\mathbb{R}$ are the clock skew and clock offset of node
$i$. In actuality, the clock skew ($\omega_i$)
and clock offset ($\phi_i$) are time varying, but we
assume they remain constant during the estimation process, which
is a reasonable assumption \cite{wu11}. The clock skew and clock offset parameters for all $N$
nodes are represented by $\bomega=[\omega_1, \omega_2,
\hdots, \omega_N]^T \in \mathbb{R}^{N \times 1}_+$ and
$\bphi=[\phi_1, \phi_2, \hdots, \phi_N]^T \in \mathbb{R}^{N
\times 1}$ respectively. Alternatively, the $2$nd part of (\ref{eq:affineTime}) shows the translation from local time $t_i$ to
the global time $t$, where $[\alpha_i, \beta_i] \triangleq [\omega^{-1}_i,\ -\phi_i\omega^{-1}_i] $ are the
calibration parameters needed to correct the local clock at node $i$.  Note that for an ideal clock,
$[\omega_i,\phi_i]=[1,0]$ subsequently implies $[\alpha_i,\beta_i]=[1,0]$ and vice versa. Following immediately, for all $N$ nodes in the
network, we have $\balpha, \bbeta \in \mathbb{R}^{N \times 1}$,
\begin{equation}
\label{eq:clockBasis}
\bomega   \triangleq \b1_{N}  \oslash \balpha  \quad\quad
\bphi     \triangleq -\bbeta  \oslash \balpha
\end{equation} \subsection{Range}\label{sec:range} In addition to clock variations, the nodes are also in motion with respect to each other. Traditionally, when the nodes are fixed \cite{patwari05} \cite{rajanCAMSAP11}, the pairwise propagation delay $\tau_{ij}$ between a node pair
$(i,j)$ is $\tau_{ij} = c^{-1}d_{ij}$, where $d_{ij}$ is the fixed distance between the node pair and $c$ is the speed of the electromagnetic wave in the medium. When the nodes are mobile, then the relative distances between the nodes are a non-linear function of time. For a node pair $(i,j)$ the propagation delay $\tau_{ij}(t) \equiv \tau_{ji}(t)$ is then, extending the affine range model \cite{rajanICASSP12}, modeled as a second order function in $t$ given by \begin{equation}
\label{eq:rangeDefinition}
\tau_{ij}(t)  = c^{-1} d_{ij}(t)= c^{-1} (\ddot{r}_{ij}t^2 + \dot{r}_{ij}t + r_{ij})
\end{equation} where $r_{ij},\dot{r}_{ij},\ddot{r}_{ij} \in \mathbb{R}$ are the range, range rate and the rate of range rate
between the node pair $(i,j)$ respectively. Substituting the equation of ideal \emph{true} time $t$
from (\ref{eq:affineTime}), we have the propagation delay $\tau_{ij}(t_i)$
in terms of the local time $t_i$, for a small duration of measurement time as
\begin{equation}
\label{eq:rangeTranslation}
\tau_{ij}(t_i) = \gamma_{ij}t_i^2 + \delta_{ij}t_i +\epsilon_{ij}
\end{equation} where $\gamma_{ij}=c^{-1} \alpha_i^2\ddot{r}_{ij}$, $\delta_{ij}   = c^{-1}(2\alpha_i^{-1}\beta_i\ddot{r}_{ij} +\alpha_i\dot{r}_{ij})$ and $\epsilon_{ij} = c^{-1}(\beta_i^2\ddot{r}_{ij} + \beta_i\dot{r}_{ij} +r_{ij})$ are the derived range parameters which incorporate the clock discrepancy of node $i$. If node $i$ is the reference
node \ie $t=t_i$, then $\gamma_{ij}=\ddot{r}_{ij},\ \delta_{ij}=\dot{r}_{ij},\ \epsilon_{ij}=r_{ij}$ as expected.
For the entire network, all $M=\begin{pmatrix} N \\ 2 \end{pmatrix}$ \emph{unique} pairwise ranges between
$N$ nodes are stacked in the vector $\br = \{r_{ij},\, \forall\ i,j = 1,2, \hdots, N;\ i < j\}\in
\mathbb{R}^{M \times 1}$ and in similar lines the
relative rate of range rates $\ddot{\br} \in \mathbb{R}^{M \times 1}$ and the
relative range rates $\dot{\br} \in \mathbb{R}^{M \times 1}$. The
derived range parameters $\bgamma, \bdelta, \bepsilon \in \mathbb{R}^{M \times 1}$ are then, $\bgamma   \triangleq c^{-1} \{\alpha_i^2\ddot{r}_{ij}\}$, $\bdelta \triangleq c^{-1}\{2\alpha_i^{-1}\beta_i\ddot{r}_{ij} +\alpha_i\dot{r}_{ij}\}$ and
$\bepsilon \triangleq c^{-1}\{\beta_i^2\ddot{r}_{ij} + \beta_i\dot{r}_{ij} +r_{ij}\}$ or alternatively, the range parameters are
\begin{subequations}\label{eq:range}
  \begin{eqnarray}
    \brrdot   &\triangleq& c\{\alpha_i^{-2}\gamma_{ij} \}   \in \mathbb{R}^{M \times 1} \label{subeq:accGamma}\\
    \brdot    &\triangleq& c\{\alpha_i^{-1}(\delta_{ij} -2\alpha^{-1}_i\beta_i\gamma_{ij})\}  \in \mathbb{R}^{M \times 1} \label{subeq:velDelta} \\
    \br       &\triangleq& c\{\epsilon_{ij} - \alpha_i^{-1}\beta_i\delta_{ij} + (\alpha_i^{-1}\beta_i)^2\gamma_{ij} \} \in \mathbb{R}^{M \times 1}\label{subeq:distEpsilon}
  \end{eqnarray}
\end{subequations}  The derived network parameters $\btheta=[\balpha, \bbeta, \bgamma, \bdelta, \bepsilon] \in \mathbb{R}^{L \times 1}$ where $L=2N+3M$, are uniquely related to the unknown clock and range parameters $\bEta= [\bomega, \bphi, \brrdot, \brdot, \br] \in \mathbb{R}^{L \times 1} $. In this paper, we intend to estimate the derived network parameters $\btheta$, given an arbitrary clock reference and communication between nodes. With $\btheta$ known, the unknown parameters $\bEta$ containing the clock parameters ($\bomega, \bphi$) and  relative range coefficients ($\br, \brdot, \brrdot$) of the network nodes can be obtained via (\ref{eq:clockBasis}) and (\ref{eq:range}). Consequently, using the range coefficients an approximate estimate of the distance over a period of time can be obtained from (\ref{eq:rangeDefinition}).

\begin{figure}[tp] \centering
\includegraphics[scale=0.21]{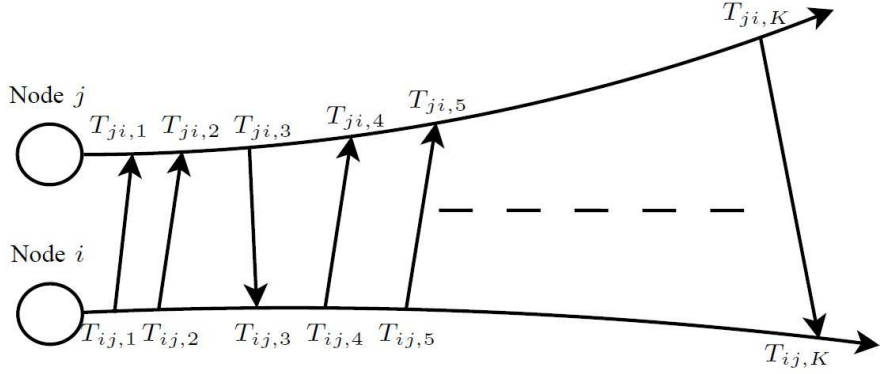}
\caption{\small{Communication between a pair of mobile nodes where the nodes transmit and receive, during which $K$ time stamps are recorded at respective nodes. Similar to \cite{rajanICASSP12,rajanCAMSAP11}, the presented model puts no pre-requisite on the sequence or number of two way communications.}}
\label{fig:figPairWise}
\end{figure} \section{Joint non-linear ranging and affine synchronization}  \label{sec:Pairwise} \label{sec:Datamodel} Consider a pair of mobile nodes $(i,j)$ such that $\{i,j\}\le N$ and $i < j$, which communicate messages back and forth, as shown
in \figurename\ \ref{fig:figPairWise}. The $k$th time stamp recorded at node $i$ when communicating with node $j$ is denoted by $T_{ij,k}$ and similarly at node $j$ the time stamp is $T_{ji,k}$. The direction of the
communication is indicated by $E_{ij,k}$, where $E_{ij,k} = +1$ for transmission from node $i$ to node $j$ and $E_{ij,k} = -1$ for
transmission from node $j$ to node $i$. In all there are $K$ time stamps recorded at each node, during which the propagation delay between the two nodes is governed by the non-linear range model given by ($\ref{eq:rangeTranslation}$). Under ideal circumstances, when the nodes are completely synchronized, the noise free $k$th communication time markers are related as $T_{ji,k}  = T_{ij,k} + E_{ij,k}\tau_{ij}(t)$
 where $E_{ij,k}= -E_{ji,k}= \pm 1$ represents the direction information of the data packet. In reality, due to measurement noise and clock uncertainties modeled in ($\ref{eq:affineTime}$) we have, $\alpha_j(T_{ji,k}+q_{j,k})+\beta_j=\alpha_i(T_{ij,k}+q_{i,k}) +\beta_i + E_{ij,k} \tau_{ij}(t_i)$,
\begin{figure*} [!b] \hrulefill \normalsize
\setcounter{eqnCounter2}{\value{equation}}
\begin{eqnarray}
\label{eq:basis1}
\alpha_iT_{ij,k} -\alpha_jT_{ji,k} +\beta_i -\beta_j +E_{ij,k}(\gamma_{ij}(T_{ij,k}+q_{i,k})^2 + \delta_{ij}(T_{ij,k}+q_{i,k} ) +\epsilon_{ij})  &=& \alpha_jq_{j,k} - \alpha_iq_{i,k} \\
\label{eq:basis2}
\underbrace{\alpha_iT_{ij,k} -\alpha_jT_{ji,k} +\beta_i -\beta_j}_{\text{Clock parameters + Measurements}}   + \underbrace{E_{ij,k}}_{\text{Direction}} \underbrace{(\gamma_{ij}(T_{ij,k})^2 + \delta_{ij}T_{ij,k} +\epsilon_{ij})}_{\text{Range parameters + Measurements}} &=& \underbrace{q_{ij,k}}_{\text{noise}}
\end{eqnarray}
\setcounter{equation}{\value{eqnCounter2}}
\vspace*{1pt}
\end{figure*}
\addtocounter{equation}{2} where $\{q_{i,k}, q_{j,k} \} \sim \cN (0, 0.5\sigma^2)$ are Gaussian \iid noise
variables plaguing the timing measurements at respective
nodes. Rearranging the terms and incorporating the range model for $\tau_{ij}(t_i)$ from
($\ref{eq:rangeTranslation}$) as a function of local time at
node $i$ we have ($\ref{eq:basis1}$). Expanding the
equation and rearranging the terms we have ($\ref{eq:basis2}$)
where, after ignoring the higher order noise terms, the noise $q_{ij,k}=  \alpha_jq_{j,k} - (\alpha_i - E_{ij,k}(2\gamma_{ij}T_{ij,k}+ \delta_{ij}))q_{i,k}$ which is modeled as $q_{ij,k}  \sim\ \cN (0, 0.5\sigma^2(\alpha_j^2 + (\alpha_i+ 2\gamma_{ij}E_{ij,k}T_{ij,k}+ \delta_{ij})^2))$. Note that the clock skews ${\omega_i}$ in reality are very
close to $1$ with errors of the order of $10^{-4}$ \cite{ieee07}. Hence,
$\alpha_j^2 \approx\ 1\ \, \forall\ j \le N\ $
and such an approximation is satisfactory and is implicitly
employed in various literature \cite{rajanCAMSAP11, wu11}.
Secondly, for $c= 3 \times 10^8$ m/s, we observe from (\ref{eq:rangeTranslation}) that
$2\gamma_{ij}E_{ij,k}T_{ij,k} + \delta_{ij}$ is negligibly small and thus the
Gaussian noise is approximated to \begin{equation}
q_{ij,k}  \sim \cN (0, \sigma^2) \label{eq:noise_k}
\end{equation} Extending ($\ref{eq:basis2}$) for all $K$ communications,
a generalized model for a pair of nodes is \begin{equation} \label{eq:basisEEPLS}
\begin{bmatrix} \bA_1 \quad \bA_2\ \end{bmatrix}
\begin{bmatrix} \alpha_{i} \\ \alpha_{j} \\ \beta_{i} \\ \beta_{j} \\ \gamma_{ij} \\ \delta_{ij} \\ \epsilon_{ij}\end{bmatrix} = \bq_{ij}
\end{equation} where $\bA_1=  \begin{bmatrix} \bt_{ij} & -\bt_{ji}  & \b1_{K}  &-\b1_{K} \end{bmatrix}$ and $ \bA_2=  \be_{ij}\odot   \begin{bmatrix} \bt^{\odot 2}_{ij}    & \bt_{ij} & \b1_K \end{bmatrix}$ contain the observation vectors \begin{eqnarray}
\label{eq:timeVector}
\bt_{ij} &=& [T_{ij,1}, T_{ij,2}, \hdots, T_{ij,K}]^T \in \mathbb{R}^{K \times 1} \\
\label{eq:directionVector} \be_{ij} &=&
[E_{ij,1},E_{ij,2}, \hdots, E_{ij,K}]^T \in
\mathbb{R}^{K \times 1}
\end{eqnarray} The time markers recorded at node $i$ and node $j$ while communicating with each other are stored in $\bt_{ij}$ and $\bt_{ji}$ respectively and $\be_{ij}$ is a known vector indicating the transmission direction for each data packet. $\bq_{ij}$ is the uncorrelated \iid noise vector given from (\ref{eq:noise_k}) as \begin{eqnarray}
\label{eq:noise}
\bq_{ij}  &=&     [q_{ij,1}, q_{ij,2}, \hdots, q_{ij,K}]^{T} \in \mathbb{R}^{K \times 1}  \nonumber  \\
          &\sim & \cN(\bzero, \sigma^2 \bI_{K})
\end{eqnarray} A unique solution to the homogenous system (\ref{eq:basisEEPLS}) can be obtained by assuming either one of $\{\alpha_i, \alpha_j \}$ and either one of $\{\beta_i, \beta_j \}$ is known, which is in turn accomplished by choosing one of the two nodes as a clock reference. More generally, asserting one of the two nodes as the reference node, say node $i$ with $[\alpha_i, \beta_i]=[1,0]$. This gives  $\bA_{ji}\btheta_{j} = -\bt_{ij} + \bq_{ij}$ where $\bA_{ji}  = [-\bt_{ji} \quad -\b1_{K} \quad \be_{ij}\odot \bt^{\odot 2}_{ij} \quad \be_{ij}\odot \bt_{ij} \quad \be_{ij} ] \in \mathbb{R}^{K \times 5}$ and  $\btheta_{ij} = [\alpha_j \quad \beta_j \quad \gamma_{ij} \quad \delta_{ij} \quad \epsilon_{ij}]^T \in \mathbb{R}^{5 \times 1}$. The \EExtended\ Pairwise Least Squares (\EEPLS) solution is obtained by minimizing the $l_2$ norm,\begin{equation}
\label{eq:ls1} \Hat{\btheta}_{ij} = \arg \min_{\btheta_{ij}} \; \|\bA_{ji}\btheta_{ij} + \bt_{ij} \|^2_2 =\ -(\bA_{ji}^T\bA_{ji})^{-1}\bA_{ji}^{T}\bt_{ij}
\end{equation} which, similar to \cite{rajanCAMSAP11}\cite{rajanICASSP12}, has a unique solution provided the number of communications $K \ge 5$, $\be_{ij} \ne -\b1_K$ and $\be_{ij} \ne +\b1_K$. The unknown parameters $[\omega_j, \phi_j, \ddot{r}_{ij}, \dot{r}_{ij}, r_{ij}]$ can be derived from  the estimate $\hat{\btheta}_{ij}= [\hat{\alpha}_j, \hat{\beta}_j, \hat{\gamma}_{ij}, \hat{\delta}_{ij}, \hat{\epsilon}_{ij}]$ using $(\ref{eq:clockBasis})$ and $(\ref{eq:range})$.\label{sec:EGLS} Aggregating (\ref{eq:basisEEPLS}), for all pairwise links in the network, we have a linear global model of the form \begin{eqnarray} \label{eq:dataModelAGLS} \overbrace{[\bT_1 \quad \bE_1 \quad \bE_2\odot\bT^{\odot 2}_2 \quad \bE_2\odot\bT_2 \quad \bE_2 ]}^{\large{\bA}} \overbrace{\begin{bmatrix} \balpha \\ \bbeta \\ \bgamma \\ \bdelta \\ \bepsilon \end{bmatrix}}^{\btheta} = \bq  \end{eqnarray} where the matrices $\bT_1,\bT_2 \in \mathbb{R}^{M_1 \times N}$ contain timing vectors recorded at all $N$ nodes, $\bE_1 \in \mathbb{R}^{M_1 \times N}$ is a matrix of $\pm\ \b1_K$ and $\bzero_K$, $\bE_2 \in \mathbb{R}^{M_1 \times M}$, where $M_1=KM$ and the noise vector is represented as $\bq = \begin{bmatrix} \bq_{12}^T, \bq_{13}^T, \hdots, \bq_{(N-1)(N)}^T\end{bmatrix}^T \in \mathbb{R}^{M_1 \times 1}$ where each $\bq_{ij}$ is given by (\ref{eq:noise}). We assume that the noise vectors for each pairwise communication $\bq_{ij}$ are uncorrelated with one another, which may not be applicable for all communication schemes \eg broadcasting. For $N=4$, $\bT_1$, $\bE_1$,  $\bT_2$,  $\bE_2$ are of the form \begin{eqnarray}
\label{eq:defTE}
\bT_1 &=&
\begin{bmatrix}
\bt_{12}& -\bt_{21} & &  \\
\bt_{13}& & -\bt_{31}&  \\
\bt_{14}& & & -\bt_{41} \\
 & \bt_{23}& -\bt_{32}& \\
 & \bt_{24}& & -\bt_{42} \\
 & &  \bt_{34} &-\bt_{43}\\
\end{bmatrix} \nonumber \\
\bE_1 &=&
\begin{bmatrix}
+\b1_{K} & -\b1_{K} & & \\
+\b1_{K} & & -\b1_{K}&  \\
+\b1_{K} & & & -\b1_{K} \\
 & +\b1_{K}& -\b1_{K}&  \\
 & +\b1_{K}& & -\b1_{K} \\
 & & +\b1_{K} & -\b1_{K}\\
\end{bmatrix} \nonumber \\
\bT_2 &=& \diag(\bt_{12}, \bt_{13}, \bt_{14},\bt_{23}, \bt_{24}, \bt_{34}) \nonumber \\
\bE_2 &=& \diag(\be_{12}, \be_{13}, \be_{14},\be_{23}, \be_{24}, \be_{34})
\end{eqnarray} and this structure can be extended for $N \ge 4 $. More generally,
the unknown vector $\btheta \in \mathbb{R}^{L \times 1}$,
where $L= 2N+3M$, can be estimated by minimizing the cost function \begin{eqnarray}
\label{eq:costFunctionEGLS}
\min_{\btheta}   && \|\ \bA\btheta\ \| ^2 \nonumber \\
\text{s.t.} && \bC\btheta = \bd
\end{eqnarray} where $\bA$ is the (rank-deficient) matrix defined in (\ref{eq:basisEEPLS})
and $\bC \in \mathbb{R}^{M_2 \times L}$, is a known constraint matrix and $\bd\in \mathbb{R}^{M_2 \times 1}$.
Assuming the constraints are
selected such that $\begin{bmatrix} \bA \\
\bC \end{bmatrix} \in \mathbb{R}^{(M_1+M_2) \times L}$
is non singular and $\bd \ne \bzero_P$,
the solution to $(\ref{eq:costFunctionEGLS})$ is obtained by
solving the $\emph{Karush-Kuhn-Tucker}$ equations
\cite{boydConvexOptimization} and is given by \begin{equation}
\label{eq:kktEEGLS}
\begin{bmatrix}
\hat{\btheta} \\
\hat{\blambda}
\end{bmatrix} = \quad
\begin{bmatrix}
2\bA^T\bA & \bC^T \\
\bC      & \bzero_{M_2, M_2}\\
\end{bmatrix}^{-1}
\begin{bmatrix}
\bzero_{L} \\
\bd
\end{bmatrix}
\end{equation} where $\blambda \in \mathbb{R}^{M_2 \times 1}$ is the Lagrange vector. If a
random node, say node $i$ is assumed to be the clock reference then the constraint matrix is
of the form \begin{eqnarray}
\label{eq:constraintC_1}
\bC =
\left[
\begin{array}{c|c|c|c|c}
\bc_i^T       & \bzero^T_{N}  & \bzero^T_{M} & \bzero^T_{M} & \bzero^T_{M} \\
\bzero^T_{N}  & \bc_i^T       & \bzero^T_{M} & \bzero^T_{M} & \bzero^T_{M}
\end{array} \right] , \quad
\bd   = \begin{bmatrix} 1 \\ 0 \end{bmatrix}
\end{eqnarray} where $\bc_i = \begin{bmatrix} \bzero^T_{i-1} ,\ 1,\ \bzero^T_{N-i}\end{bmatrix}^T \in \mathbb{R}^{N \times 1}$. Similar to \cite{rajanCAMSAP11} \cite{rajanICASSP12}, despite missing links, network wide synchronization is still feasible using the proposed algorithms if each node has at least one link with any other node in the network. \section{Constrained Cramer Rao lower bound} \label{sec:CRB}
In order to verify the performance of the proposed algorithm, we derive a Constrained Cramer Rao lower Bound (CCRB) for the model in (\ref{eq:dataModelAGLS}), where Gaussian noise is assumed on the time markers . The CCRB on the error variance for an unbiased estimator is given by \cite{stoica1998} \begin{equation} \label{eq:crbEEGLS}
{\text{\Large{$ \varepsilon $}}} \left \{ (\hat{\btheta}-\btheta)(\hat{\btheta}-\btheta)^T \right \} \ge \cSigma_{\theta}
=\ \bU(\bU^T\bF\bU)^{-1}\bU^T
\end{equation} where $\cSigma_{\theta}$ is the lower bound on $\btheta$, $\bU \in \mathbb{R}^{L \times (L-M_2)}$ with $L=2N+3M$ is an orthonormal basis for the null space of the constraint matrix $\bC$ with $M_2$ constraints, and $\bF = \sigma^{-2}\bA^T\bA \in
\mathbb{R}^{L \times L}$ is the Fisher Information Matrix. Since the system parameters $\bEta=  [\bomega,\ \bphi,\ \brrdot,\ \brdot,\ \br ]$ can be uniquely derived from $\btheta$, we have the CRB on the estimates of $\bEta$ from standard error propagation formulas, $\cSigma_{\eta} =\  \bJ_{\theta{\eta}}\ \bSigma_{\theta}\ \bJ^T_{\theta{\eta}}$ where $\bSigma_{\theta}$ is given by ($\ref{eq:crbEEGLS}$) and $\bJ_{\theta{\eta}} \in \mathbb{R}^{L \times L} $
is the Jacobian of the transformation of $\bEta$ from $\btheta$, which is given by (\ref{eq:JacobianEEGLS}), where $\cA= \diag(\balpha)^{-1} \in \mathbb{R}^{N\times N} , \cB =\diag(\bbeta) \in \mathbb{R}^{N\times N} , \tilde{\cA}= \diag(\tilde{\balpha})^{-1} \in \mathbb{R}^{M\times N}$ and $\tilde{\cB}= \diag(\tilde{\bbeta}) \in \mathbb{R}^{M\times N}$. Furthermore for $N=4$, $\cG \in \mathbb{R}^{M\times N}, \cD \in \mathbb{R}^{M\times N}$ are of the form  \begin{eqnarray}
\label{eq:crbJacobComponents_2}
\cG &=&
\begin{bmatrix}
\gamma_{12} & \gamma_{13} & \gamma_{14} & & & \\
            &             &             & \gamma_{23} & \gamma_{24}   & \\
            &             &             &             &               & \gamma_{34} \\
            &             & \bzero^T_M  &             &               &
\end{bmatrix} \nonumber \\
\cD &=&
\begin{bmatrix}
\delta_{12} & \delta_{13} & \delta_{14} & & & \\
            &             &             & \delta_{23} & \delta_{24}   & \\
            &             &             &             &               & \delta_{34} \\
            &             & \bzero^T_M  &             &               &
\end{bmatrix}
\end{eqnarray} which can be extended for $N\ge4$ in a straightforward way. \begin{figure*} [!b] \normalsize
\vspace*{1pt}\hrulefill
\setcounter{eqnCounter2}{\value{equation}}
\begin{eqnarray}
\label{eq:JacobianEEGLS}
\bJ_{\theta{\eta}}
& \triangleq &
\begin{bmatrix}
\dfrac{\partial \bEta}{\partial\btheta^T}
\end{bmatrix}
=
\begin{bmatrix}
-\cA^2        & \cA^2\cB      & -2c\cA^3\cG    &  -c\cA^2\cD +2c\cA^3\cB\cG   & c\cA^2\cB\cD       \\
\bzero_{N,N}  & -\cA\         & \bzero_{N,M}  &  -2c\cA^2\cG                  & -c\cA\cD + 2c\cB^3\cG \\
\bzero_{M,N}  & \bzero_{M,N}  & -c\cA^2       &  -2c\tilde{\cA}^2\tilde{\cB}  & c\tilde{\cB}^2            \\
\bzero_{M,N}  & \bzero_{M,N}  & \bzero_{M,N}  &  c\tilde{\cA}                 & c\tilde{\cA}\tilde{\cB}   \\
\bzero_{M,N}  & \bzero_{M,N}  & \bzero_{M,N}  &  \bzero_{M,N}                 & c\bI_M                    \\
\end{bmatrix}
\label{eq:JacobianFormEEGLS}
\end{eqnarray}
\setcounter{equation}{\value{eqnCounter2}}
\end{figure*}
\addtocounter{equation}{1}

\section{Simulations} \label{sec:simulations}
\begin{figure}[htp]
\centering
  \includegraphics[scale=0.6]{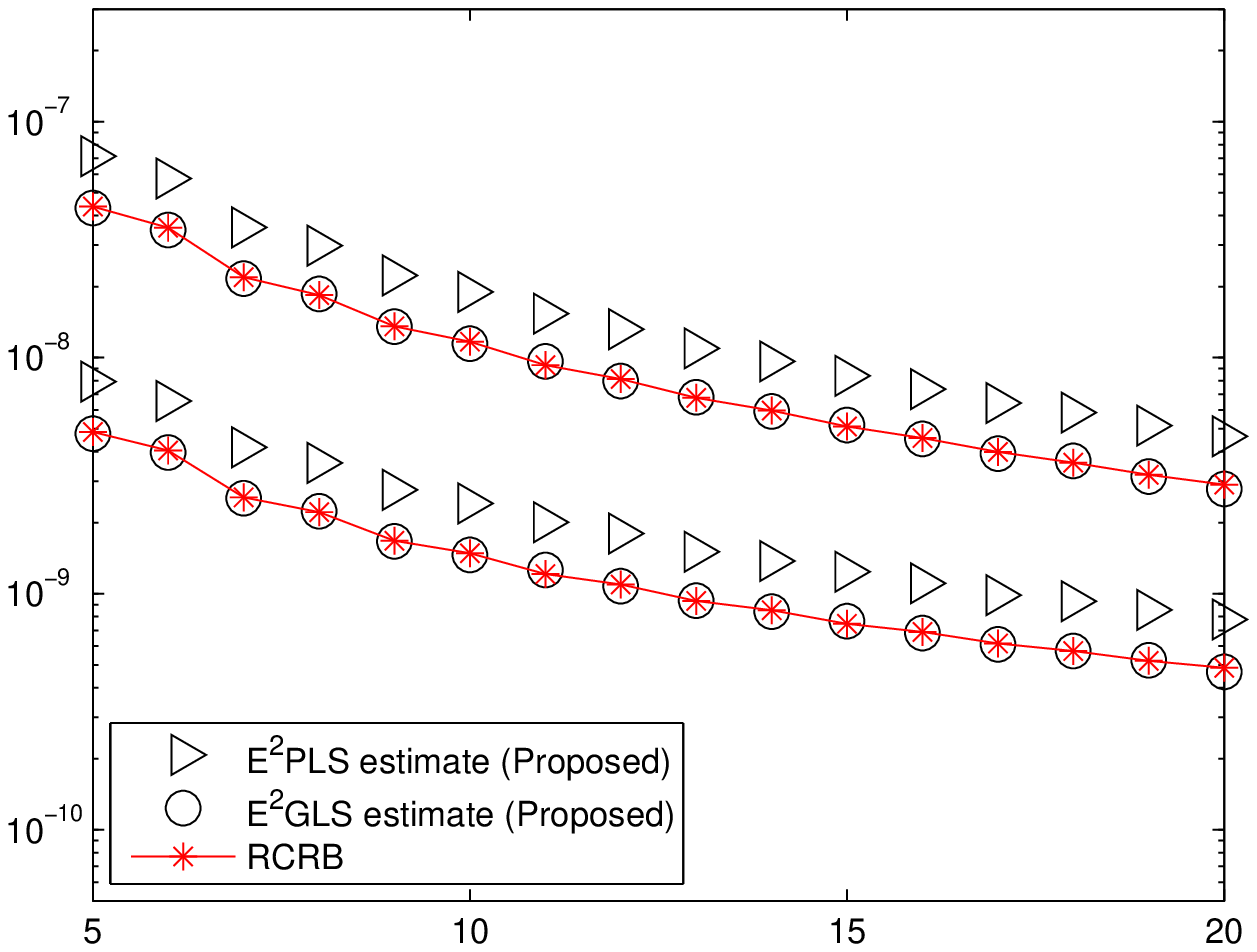}
  \rput*{90}(+0.2,9.8){\scriptsize{RMSE of clock parameters ($\hat{\bomega}$, $\hat{\bphi}$)}}
  \rput(+4.2, 11.5){\scriptsize{Clock skews ($\bomega$)}}
  \rput(+4.2, 10.2){\scriptsize{Clock offsets ($\bphi$)}}
\centering
  \includegraphics[scale=0.6]{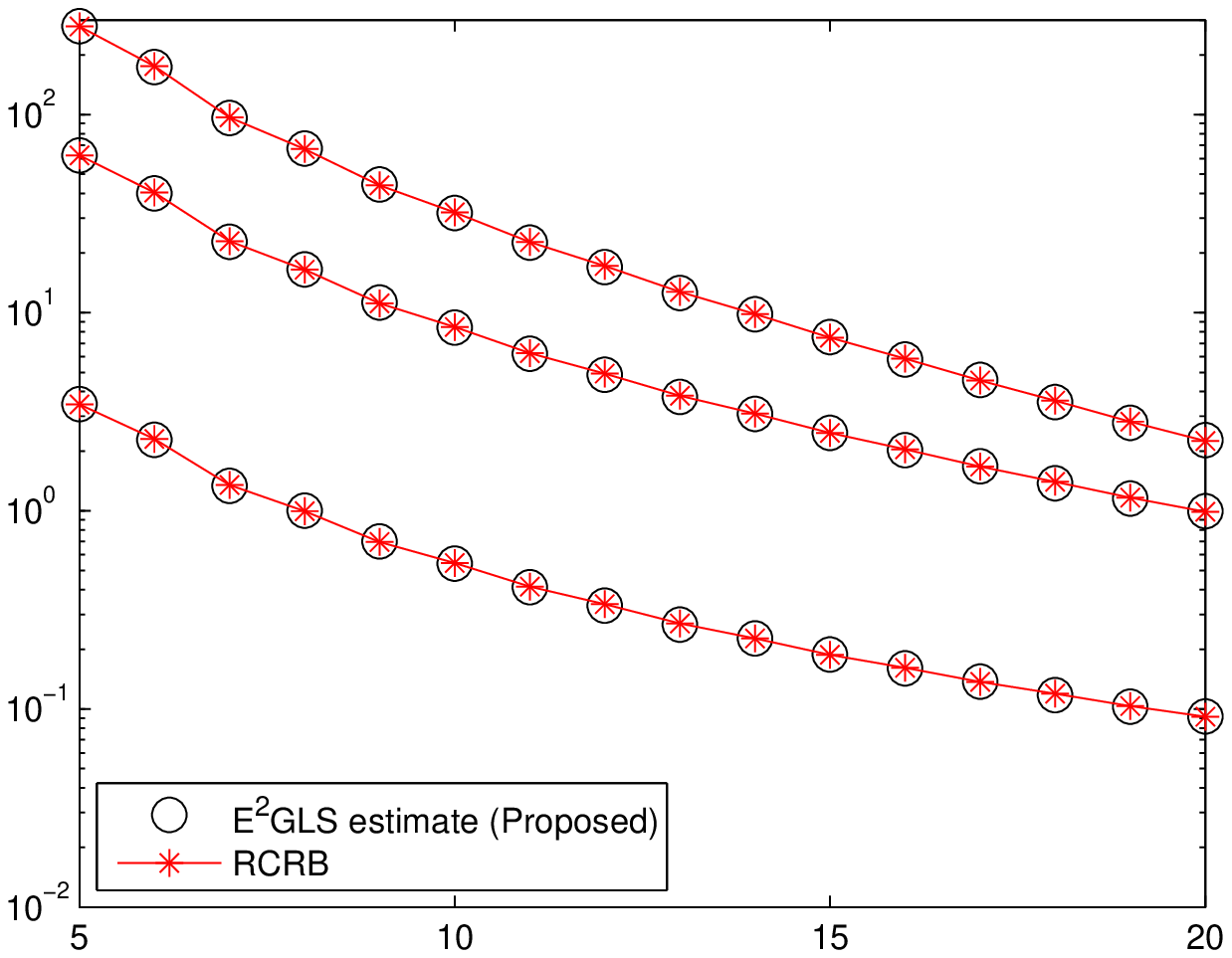}
  \rput(0.5,   0.7){\scriptsize{Number of two way communications (K)}}
  \rput*{90}(-4,3.8){\scriptsize{RMSE of range parameters ($\hat{\brrdot}$, $\hat{\brdot}$, $\hat{\br}$)}}
  \rput(-0.0, 2.5){\scriptsize{Rate of range rates ($\brrdot$)}}
  \rput(-0.0, 3.8){\scriptsize{Range rates     ($\brdot$)}}
  \rput(-0.0, 4.8){\scriptsize{Ranges        ($\br$)}}
\caption{\small{Root Mean Square Error (RMSE) and RCRB plots of (a) estimated clock parameters $[\hat{\bomega}, \hat{\bphi}]$
  and (b) range parameters $[\hat{\brrdot}$, $\hat{\brdot}$, $\hat{\br}]$ for a network of $N=4$ nodes, where the noise is Gaussian with $\sigma=0.01 \mu s$}}
\rput(0,9.3){(a)}
\rput(0,2.2){(b)}
\label{fig:Results}
\end{figure}

We consider a network of $4$ nodes, each capable of two way communication with each other. The clock skews ($\bomega$) and clock offsets ($\bphi$) and  of the nodes are uniform randomly distributed in the range $[1-10 \text{ppm},1+10\text{ppm}]$ and $[-10,+10]$ seconds respectively. The range parameters ($\brrdot, \brdot, \br$) of the nodes are uniformly distributed in the range $[-0.1, +0.1]$ m/$\text{s}^2$ , $[-1, +1]$ m/s and $(0, 10]$ Km respectively, which is acceptable for satellites in (selective) orbits around the moon \cite{montenbruck00}, for short intervals of time. The transmission time markers $\bt_{ij}$ are linearly distributed between $0.1$ to $10$ seconds, for a number of two way communication links $K$ spanning from $5$ to $20$, wherein the nodes transmit and receive time stamps alternatingly\cite{rajanCAMSAP11}. The metric used to evaluate the performance of the estimators is the Root Mean Square Error (RMSE) and without loss of generality, node $1$ is considered to be the reference node with $[\alpha_1\ \beta_1]=[1, 0]$ and the Gaussian noise on the time markers has a standard deviation $\sigma=0.01 \mu s$. Furthermore, along with the RMSE plots, the Root mean square of the Constrained Cramer Rao Bound (RCRB) derived in Section \ref{sec:CRB} are also plotted. The \EEPLS\ algorithm is independently applied, pairwise from node $1$ to every other node to estimate all the unknown clock parameters ($\bomega, \bphi$) and for the entire network, the \EEGLS\ algorithm is applied. \figurename\ \ref{fig:Results}(a) shows the RMSE plots vs the number of communications $K$ for the clock skew $(\bomega)$ and the clock offset $(\bphi)$. The \EEGLS\ estimate outperforms the \EEPLS\ estimate for clock parameter estimation, which is expected, since the total number of communication links available for the \EEGLS\ estimate is greater than that for \EEPLS\ \ie $M > (N-1)$ for $N\ge2$. The RMSE of the relative range parameters $[\brrdot, \brdot, \br]$ are plotted in \figurename\ \ref{fig:Results}(b) and all the estimates, perhaps not surprisingly,  achieve the RCRB derived in (\ref{eq:crbEEGLS}) asymptotically.

\section{Conclusions} \emph{For a cluster of model nodes, the pairwise distances between the nodes are always non-linear and hence a second order range model in conjunction with an affine clock model is proposed}. The \EEPLS\ and \EEGLS\ algorithms are least squares solutions for estimating the clock ($\bomega, \bphi$) and range parameters ($[\brrdot, \brdot, \br]$), for a pair of nodes and the entire network respectively. Given these parameters, the nodes can be synchronized to a reference clock and the time varying pairwise distance estimated approximately. A new Constrained Cramer Rao Bound (CCRB) is derived and the proposed estimators approach the theoretical limits asymptotically. 

\bibliographystyle{core/IEEEtran}
\bibliography{core/myRef,strings,refs}
\end{document}